\documentstyle[aps,12pt]{revtex}
\begin{document}
\title{{\bf Radiation From a Sine-Gordon Soliton Propagating in an External
Potential}}
\author{{\bf A\ Calogeracos}\thanks{acal@hol.gr, acal@boojum.fi}}
\address{Low Temperature Laboratory, Helsinki University of Technology, Otkaari 3A,\\
02150, Espoo, Finland, and} 
\address {Centre for Theoretical Physics, University of Sussex, Brighton, BN1 9QJ, 
UK\thanks{permanent address\
NCA\ Research Consultants, PO Box 61147,Maroussi 151 22, Greece} }
\date{SUSX-TH-99-031}
\maketitle
\bigskip
\bigskip
\begin{abstract}
\noindent The interaction of a fast moving sine-Gordon soliton with an
external periodic potential is examined. The resulting equation of motion
for the collective coordinate representing the position of the soliton is
given in relativistic form. We examine the radiation emitted due to the
interaction of the soliton with the potential and we calculate the potential
dependent part of the time evolution equation for the creation and
annihilation operators for fluctuations.
\end{abstract}

\newpage

\section{Introduction.}

\-In this note we consider a sine-Gordon soliton in the presence of a
periodic potential. The soliton is fast moving, thus relativistic effects
shall be taken into account. In particular we expect the total energy of the
system to be divided into two contributions, one equal to $M/\sqrt{1-B^{2}}$
corresponding to a soliton of mass $M$ moving at velocity $B$ and another
pertaining to the field of fluctuations about the soliton solution. A
similar separation should hold for the total momentum of the system where
the term $MB/\sqrt{1-B^{2}}$ is expected to appear. Clearly expanding around
the usual static solution (\ref{1}) is certainly not good enough:\ A fast
moving soliton suffers Lorentz contraction and the Lorentz factor should
appear in the classical solution. \- The soliton interacts both with the
fluctuating field and with the external potential; it will prove both
natural and fruitful to treat the velocity $B$ as a dynamical variable. This
is essential if recoil effects are to be taken into account. A formalism
along these lines is presented in \cite{1}; however the present note is
meant to be self-contained. Note that (a) the model considered in \cite{1}
was a pure $\phi ^{4}$ theory with no external potential (the work focused
on the interaction of the soliton with the fluctuations), (b) the emphasis
was on the quantization of the theory. A model of a soliton in an external
potential is presented in \cite{2} but there the aim was to examine the
motion of the soliton, radiation effects being disregarded. The problem
examined in the present note has been looked at in \cite{4}; here we put an
emphasis on a relativistic treatment of the problem.

\section{\noindent Basics.}

The simple sine-Gordon Lagrangean is given by (for a review see \cite{3}, 
\cite{8}) 
\begin{equation}
L=\int dx^{\prime }\left\{ \frac{1}{2}\left( \frac{\partial \phi }{\partial
t^{\prime }}\right) ^{2}-\frac{1}{2}\left( \frac{\partial \phi }{\partial
x^{\prime }}\right) ^{2}-U_{sG}(\phi )\right\}  \label{2}
\end{equation}

\noindent where 
\begin{equation}
U_{sG}(\phi )=\frac{\alpha }{\beta ^{2}}\left( 1-\cos (\beta \phi )\right)
\label{6}
\end{equation}

\noindent (Lab frame coordinates are denoted by $(t^{\prime },x^{\prime }))$%
. We make the choice 
\begin{equation}
\alpha =\beta =1  \label{8}
\end{equation}
\noindent and then 
\begin{equation}
U_{sG}(\phi )=1-\cos \phi  \label{13}
\end{equation}
\noindent Eventually we will consider the soliton in an external perturbing
potential of the form \cite{4} 
\begin{equation}
U(\phi )=U_{sG}(\phi )\left( 1+V(x^{\prime })\right)  \label{25}
\end{equation}
\begin{equation}
V(x^{\prime })=\varepsilon \cos (k_{0}x^{\prime })  \label{v}
\end{equation}
where $\varepsilon $ is a small quantity. The equation of motion resulting
from (\ref{2}) is 
\begin{equation}
\frac{\partial ^{2}\phi }{\partial t^{\prime 2}}-\frac{\partial ^{2}\phi }{%
\partial x^{\prime 2}}+U_{sG}^{\prime }(\phi )=0  \label{4}
\end{equation}

\noindent The static version 
\begin{equation}
-\frac{d^{2}\phi }{dx^{\prime 2}}+U_{sG}^{\prime }(\phi )=0  \label{4b}
\end{equation}

\noindent admits (amongst others) the solution 
\begin{equation}
\Phi _{c}(x^{\prime })=4\arctan (e^{x^{\prime }})  \label{1}
\end{equation}

\noindent where $\Phi _{c}\rightarrow 0$ for $x^{\prime }\rightarrow -\infty 
$ and $\Phi _{c}\rightarrow 2\pi $ for $x^{\prime }\rightarrow \infty $. The
mass of the soliton is quite generally given by 
\begin{equation}
M=\int dx^{\prime }\left\{ \frac{1}{2}\left( \frac{\partial \Phi _{c}}{%
\partial x^{\prime }}\right) ^{2}+U_{sG}(\Phi _{c})\right\} =\int dx^{\prime
}\left( \frac{\partial \Phi _{c}}{\partial x^{\prime }}\right) ^{2}
\label{19}
\end{equation}

\noindent and for the particular choice (\ref{8}) $M=8.$ Fluctuations around
the classical solution satisfy the equation (expand (\ref{4}) around $\Phi
_{c}$) 
\begin{equation}
-\frac{d^{2}f}{dx^{\prime 2}}+U_{sG}^{\prime \prime }(\Phi _{c})f=\omega
^{2}f  \label{5}
\end{equation}

\noindent with $\omega $ the corresponding frequency. Note that the space
derivative of the kink 
\begin{equation}
\Phi _{c}^{\prime }(x^{\prime })=\frac{2}{\cosh x^{\prime }}  \label{zero}
\end{equation}
\noindent is a solution corresponding to zero frequency (the zero mode) (in
the last equation we used (\ref{1})). This simply reflects the fact that a
translated kink $\Phi _{c}(x^{\prime }+a)$ is also a static solution. The
normal modes corresponding to a wavevector $k$ (not to be confused with $%
k_{0}$ of (\ref{v}) that is a characteristic of the potential) are given in 
\cite{4} : 
\begin{equation}
\frac{1}{\omega (k)\sqrt{2\pi }}e^{ikx^{\prime }}\left( k+i\tanh x^{\prime
}\right)  \label{14}
\end{equation}

\noindent where $\omega =\sqrt{k^{2}+\mu ^{2}}$ , and $\mu =1$ with the
choice (\ref{8}). We shall write the normal modes in the form 
\begin{equation}
f(k,x)=\tilde{A}(k,x^{\prime })e^{ikx^{\prime }}  \label{14b}
\end{equation}

\noindent where the prefactor $\tilde{A}(k,x^{\prime })$ can be read off (%
\ref{14}) modulo a norming factor; we will work with energy normalized
wavefunctions 
\[
\int dxf^{*}(k_{1},x)f(k_{2},x)=\delta (\omega _{1}-\omega _{2}) 
\]

The Hamiltonian is 
\begin{equation}
H=\int dx\left\{ \frac{1}{2}\left( \frac{\partial \phi _{c}}{\partial
x^{\prime }}\right) ^{2}+\frac{\pi _{\phi }^{2}}{2}+U_{sG}(\phi )\right\}
\label{15}
\end{equation}

\noindent The fields are expanded in terms of the zero mode and the
travelling modes 
\begin{equation}
\phi (x^{\prime },t^{\prime })=\Phi _{c}(x^{\prime })+q_{z}\Phi _{c}^{\prime
}(x^{\prime })+\frac{1}{\sqrt{4\pi }}\int \frac{d\omega }{\sqrt{\omega }}%
\left\{ \tilde{a}^{\dagger }(\omega )f(k,x^{\prime })e^{i\omega t^{\prime }}+%
\tilde{a}(\omega )f^{*}(k,x^{\prime })e^{-i\omega t^{\prime }}\right\}
\label{16}
\end{equation}

\begin{equation}
\pi _{\phi }(x^{\prime },t^{\prime })=\frac{p_{z}}{M}\Phi _{c}^{\prime
}(x^{\prime })+\frac{1}{\sqrt{4\pi }}\int d\omega \sqrt{\omega }\left\{ 
\tilde{a}^{\dagger }(\omega )f(k,x^{\prime })e^{i\omega t^{\prime }}-\tilde{a%
}(\omega )f^{*}(k,x^{\prime })e^{-i\omega t^{\prime }}\right\}  \label{17}
\end{equation}

\noindent with the commutation relations 
\begin{equation}
\left[ q_{z},p_{z}\right] =1  \label{18}
\end{equation}

\[
\left[ \tilde{a}(\omega ),\tilde{a}^{\dagger }(\omega )\right] =\delta
(\omega -\omega ^{\prime }) 
\]

\noindent $\phi ,\Pi _{\phi }$ obey the standard canonical commutation
relations. (NB\ In what follows and for the sake of brevity integration over 
$\omega $ will also stand for summation over the two directions of
incidence.)

Notice that $p_{z}$ appears in the Hamiltonian in the form $%
%TCIMACRO{\dfrac{p_{z}^{2}}{2M}}
%BeginExpansion
{\displaystyle {p_{z}^{2} \over 2M}}
%EndExpansion
$ through the $%
%TCIMACRO{\dfrac{\pi _{\phi }^{2}}{2}}
%BeginExpansion
{\displaystyle {\pi _{\phi }^{2} \over 2}}
%EndExpansion
$ term. Thus $p_{z}$ can be identified as the momentum of the soliton. The
fact that the kinetic energy of the soliton appears in the nonrelativistic
form is connected to the fact that we consider fluctuations about the
zero-velocity solution (\ref{1}). Clearly the inclusion of relativistic
effects requires expansion about a boosted version of (\ref{1}). Neglecting
the interaction with fluctuations the motion of the kink at a
nonrelativistic speed $u$ corresponds to 
\begin{equation}
q_{z}=ut^{\prime }  \label{30}
\end{equation}
\begin{equation}
p_{z}=Mu  \label{31}
\end{equation}

\noindent compatible with $\pi _{\phi }=\partial \phi /\partial t^{\prime }.$

We introduce the fluctuating field 
\begin{equation}
\chi (x^{\prime },t^{\prime })\equiv \phi (x^{\prime },t^{\prime })-\Phi
_{c}(x^{\prime })  \label{20}
\end{equation}

\noindent and split the Hamiltonian (\ref{15}) into a part quadratic in $%
\chi $ and another part involving higher order terms. Then the quadratic
part (diagonalized by the normal modes) does not involve $q_{z}$. Thus the
momentum of the kink (identified with $p_{z}$) is conserved; it can change
only through interactions either with the fluctuating field or with an
external potential.

\section{\noindent \noindent The moving soliton.}

\subsection{The Hamiltonian formalism}

In order to treat a fast moving soliton we employ the strategy used in \cite
{1} and in a somewhat different context in \cite{5}, namely we go over to a
non-inertial frame comoving with the soliton. The definition of the position
of the soliton is of course not a straightforward affair and it will occupy
us later on. The new coordinates are defined by 
\begin{equation}
x^{\prime }=x+X(t)\text{ , }t^{\prime }=t\text{ , }B\equiv \dot{X}(t)
\label{21}
\end{equation}

\noindent We redefine the field so that 
\begin{equation}
\Phi (x)=\phi (x+X)  \label{24}
\end{equation}
\noindent (Transformations (\ref{21}), (\ref{24}) coincide with the
translation transformation implemented quantum-mechanically in \cite{6},
section 3.1). Thus 
\begin{equation}
\frac{\partial \phi }{\partial x^{\prime }}=\frac{\partial \Phi }{\partial x}%
\text{ , }\frac{\partial \phi }{\partial t^{\prime }}=\frac{\partial \Phi }{%
\partial t}-B\frac{\partial \Phi }{\partial x}  \label{22}
\end{equation}

\noindent

Lagrangean (\ref{2}) with the potential (\ref{25}) can be readily written in
terms of the new coordinates 
\begin{equation}
L=\int dx\left\{ \frac{1}{2}\left( \frac{\partial \Phi }{\partial t}-B\frac{%
\partial \Phi }{\partial x}\right) ^{2}-\frac{1}{2}\left( \frac{\partial
\Phi }{\partial x}\right) ^{2}-U_{sG}(\Phi (x))(1+V(x+X))\right\}
\label{230}
\end{equation}

\noindent The momentum $\Pi $ conjugate to $\Phi $ is 
\begin{equation}
\Pi =\frac{\partial \Phi }{\partial t}-B\frac{\partial \Phi }{\partial x}
\label{26}
\end{equation}

\noindent whereas the momentum $P$ conjugate to $X$ satisfies the constraint 
\begin{equation}
P+\int dx\Pi \frac{\partial \Phi }{\partial x}=0  \label{27}
\end{equation}

\noindent The constraint reflects the arbitrariness in what we mean by $X$
(i.e. the freedom to perform coordinate transformations). The canonical
Hamiltonian $H_{c}=p\dot{q}-L$ is 
\begin{equation}
H_{c}=\int dx\left\{ \frac{\Pi ^{2}}{2}+\frac{1}{2}\left( \frac{\partial
\Phi }{\partial x}\right) ^{2}+U_{sG}(\Phi (x))(1+V(x+X))\right\}  \label{28}
\end{equation}

\noindent The numerical value of $H_{c}$ coincides with the total energy of
the system.

The total Hamiltonian (i.e. the quantity that yields the equations of
motion) consists of the canonical Hamiltonian $H_{c}$ plus the constraint
multiplied by an arbitrary Lagrange multiplier \cite{7} 
\begin{equation}
H_{c}+B\left( P+\int dx\Pi \frac{\partial \Phi }{\partial x}\right)
\label{29}
\end{equation}

\noindent The identification of the Lagrange multiplier $B$ as the velocity
follows immediately by commuting $X$ with (\ref{29}) and taking into account
the fact that $X$ and $P$ are canonically conjugate.

Since the velocity changes as a result of interactions with the fluctuations
and/or the external potential it is natural to elevate it to the status of a
dynamical variable and introduce its conjugate momentum $p_{B}$%
\begin{equation}
\left[ B,p_{B}\right] =1  \label{33}
\end{equation}

\noindent We take $B,p_{B}$ to commute with all other variables $\Phi ,\Pi
,X,P$. We require $p_{B}$ to vanish as a constraint and introduce the
(final) total Hamiltonian 
\begin{equation}
H=H_{c}+B\left( P+\int dx\Pi \frac{\partial \Phi }{\partial x}\right) +ap_{B}
\label{32}
\end{equation}

\noindent where we added to the Hamiltonian (\ref{29}) constraint $p_{B}$
multiplied by a Lagrange multiplier $a$. Commuting $B$ with $H$ we get 
\begin{equation}
\dot{B}=a  \label{34}
\end{equation}

\noindent Thus $a$ is identified as the acceleration of the kink. Requiring $%
p_{B}$ to be conserved we end up with the original constraint (\ref{27}).

The situation is reminiscent of what happens in electrodynamics. The
momentum conjugate to the scalar potential (the latter being the analog of $%
B $ in the present case) vanishes and the requirement that the constraint be
conserved leads to Gauss's law (the analog of (\ref{27})) as a secondary
constraint. The gauge invariance reflects the freedom of performing
coordinate transformations of the type (\ref{21}). It will be lifted when we
impose a subsidiary condition (thus implicitly defining the position of the
soliton and hence the meaning of a comoving frame).

If we neglect for the moment the effect of the perturbation $V$, the field
equation resulting from (\ref{230}) (or (\ref{28}) and (\ref{32})) for
constant velocity $B$ is 
\begin{equation}
\frac{\partial ^{2}\Phi }{\partial t^{2}}-\frac{\partial ^{2}\Phi }{\partial
x^{2}}+U_{sG}^{\prime }(\Phi )-2B\frac{\partial ^{2}\Phi }{\partial
x\partial t}+B^{2}\frac{\partial ^{2}\Phi }{\partial x^{2}}=0  \label{351}
\end{equation}

\noindent A $t$ independent solution is trivially found to be 
\begin{equation}
\Phi _{c}\left( \frac{x}{\sqrt{1-B^{2}}}\right)  \label{352}
\end{equation}

\noindent \noindent where $\Phi _{c}$ is the same as (\ref{1}) (i.e. the
space variable is simply rescaled by the Lorentz factor). It is natural to
expand the fields $\Phi ,\Pi $ about the classical solution (\ref{352})
introducing the fluctuating fields $\chi $ and $\pi $ (the latter not to be
confused with $\pi _{\phi }$ of section 2). (NB\ in what follows $\Phi
_{c}^{^{\prime }}$ will refer to the derivative of $\Phi _{c}$ with respect
to its argument and not just with respect to $x.)$ 
\begin{equation}
\Phi (x,t)=\Phi _{c}\left( \frac{x}{\sqrt{1-B^{2}}}\right) +\chi (x,t)
\label{35}
\end{equation}

\begin{equation}
\Pi (x,t)=-\frac{B}{\sqrt{1-B^{2}}}\Phi _{c}^{^{\prime }}\left( \frac{x}{%
\sqrt{1-B^{2}}}\right) +\pi (x,t)  \label{36}
\end{equation}

\noindent The existence of the first term in (\ref{36}) is a consequence of
the $\partial /\partial x$ term in (\ref{26}). It follows from (\ref{35}), (%
\ref{36}) that $\chi ,\pi $ satisfy canonical commutation relations 
\begin{equation}
\left[ \chi (x,t),\pi (x^{\prime },t)\right] =1  \label{39}
\end{equation}

\noindent It is important however to realize that $p_{B}$ does not commute
with $\chi $ and $\pi $. Given that $p_{B}$ commutes with $\Phi ,\Pi $ it
follows immediately from (\ref{35}), (\ref{36}) that 
\begin{equation}
\left[ \chi (x,t),p_{B}\right] =-\Phi _{c}^{\prime }\left( \frac{x}{\sqrt{%
1-B^{2}}}\right) \frac{d}{dB}\left( \frac{1}{\sqrt{1-B^{2}}}\right)
\label{37}
\end{equation}

\begin{equation}
\left[ \pi (x,t),p_{B}\right] =\Phi _{c}^{^{\prime }}\left( \frac{x}{\sqrt{%
1-B^{2}}}\right) \frac{d}{dB}\left( \frac{B}{\sqrt{1-B^{2}}}\right) +\frac{%
B^{2}}{\left( 1-B^{2}\right) ^{2}}\Phi _{c}^{^{\prime \prime }}\left( \frac{x%
}{\sqrt{1-B^{2}}}\right)  \label{38}
\end{equation}

We turn to the field momentum and express it in terms of the static solution
and the fluctuations (and use (\ref{19})): 
\begin{equation}
\int dx\Pi \frac{\partial \Phi }{\partial x}=-\frac{MB}{\sqrt{1-B^{2}}}+\int
dx\pi \frac{\partial \chi }{\partial x}+\frac{1}{\sqrt{1-B^{2}}}\int
dx\left\{ \Phi _{c}^{^{\prime }}\pi +\frac{B}{\sqrt{1-B^{2}}}\Phi
_{c}^{^{\prime \prime }}\chi \right\}  \label{41}
\end{equation}

\noindent It is gratifying that in the absence of fluctuations we get the
expected relativistic expression for the kink's momentum.

\subsection{The subsidiary condition.}

We have to define what we mean by the position of the soliton or in other
words give a definition of the comoving frame. This is done through the
imposition of a constraint (a subsidiary condition) that lifts the freedom
under coordinate transformations and assigns physical meaning to the
variable $X.$ The choice of the suitable constraint is dictated by three
criteria each one having its own physical motivation. It is not obvious at
the outset that all three can be simultaneously satisfied. The fact that
they can adds physical appeal to the separation of the total field $\Phi $
to a soliton part and a fluctuating part.

The first criterion stipulates that in the presence of fluctuations the
total momentum splits neatly to the soliton and fluctuation momenta given
respectively by minus the first and second terms in (\ref{41}) and that the
term in braces vanishes. In other words the constraint that lifts the gauge
invariance is 
\begin{equation}
C\equiv \int dx\left\{ \Phi _{c}^{^{\prime }}\pi +\frac{B}{\sqrt{1-B^{2}}}%
\Phi _{c}^{^{\prime \prime }}\chi \right\} =0  \label{420}
\end{equation}

\noindent The fact that $C$ does not contribute numerically in (\ref{41})
does not mean that it should be discarded; it may well contribute to the
equations of motion since the LHS\ of (\ref{41}) appears in the total
Hamiltonian (\ref{32}). It turns out that it does not contribute; see the
remark following (\ref{400e}).

As a second criterion we require that the $energy$ $H_{c}$ of the system
does not contain any terms linear in the fluctuations. In other words that
the same state of affairs as in the case of the total momentum prevails. It
turns out that this requirement is satisfied; see (\ref{400e}) and the
remark following it.

The third criterion stipulates that constraint $C$ commute with the
quadratic part $H_{0}$ (\ref{50}) of the total Hamiltonian. This motivation
is linked to the acceleration of the kink. The acceleration will be
determined in the next subsection (eqn (\ref{451a})) by requiring that
constraint $C$ commute with the total Hamiltonian (\ref{32}), and it has two
physical origins:\ (i) The interaction of the kink with the fluctuations,
i.e. terms in the Hamiltonian beyond the quadratic in expansion (\ref{400c}%
); (ii) The interaction with the external potential $V$. Thus in the absence
of (i) and (ii) the acceleration ought to vanish, i.e. $C$ should commute
with $H_{0}$ (\ref{50}). This is shown in (\ref{51b}) below.

We substitute expansions (\ref{35}), (\ref{36}) in $H_{c}$ (\ref{28}) while
expanding $U_{sG}(\Phi )$ up to terms quadratic in $\chi $. Use of (\ref{19}%
) and straight algebra yields 
\begin{eqnarray}
H_{c} &=&\frac{M}{\sqrt{1-B^{2}}}+  \nonumber \\
&&+\int dx\left\{ \frac{1}{2}\left( \frac{\partial \chi }{\partial x}\right)
^{2}+\frac{\pi ^{2}}{2}+\frac{1}{2}U_{sG}^{\prime \prime }\left( \Phi
_{c}\right) \chi ^{2}+V\text{-independent higher orders in }\chi \right\} + 
\nonumber \\
&&+\left\{ V(x+X)U_{sG}\left( \Phi _{c}\right) +V\text{-dependent higher
orders in }\chi \right\} +  \label{400c} \\
&&+\int dx\left\{ -\frac{B}{\sqrt{1-B^{2}}}\Phi _{c}^{^{\prime }}\pi -\frac{1%
}{1-B^{2}}\Phi _{c}^{^{\prime \prime }}\chi +U_{sG}^{^{\prime }}\left( \Phi
_{c}\right) \chi \right\}  \nonumber
\end{eqnarray}

\noindent Thus $H_{c}$ naturally splits to a number of contributions: The
first term coincides with the covariant expression for the energy of a
particle. The next three terms are quadratic in the fluctuations and enter
in the calculation of the normal modes of the system. There also are terms
describing the interaction of the kink with the fluctuations and/or the
external potential. The last term in braces linear in fluctuations can be
written using (\ref{4b}) in the form 
\begin{equation}
-\frac{B}{\sqrt{1-B^{2}}}\int dx\left\{ \Phi _{c}^{\prime }\pi +\frac{B}{%
\sqrt{1-B^{2}}}\Phi _{c}^{\prime \prime }\chi \right\}  \label{400e}
\end{equation}

\noindent Hence the second criterion spelt out in the beginning of this
subsection is satisfied: the energy does not depend linearly on the
fluctuations. Notice that (\ref{400e}) cancels exactly with the last term in
(\ref{41}) when they are both substituted in (\ref{32}). Thus the total
Hamiltonian $H$ is written in terms of $\chi ,\pi $ in the form 
\begin{eqnarray}
H &=&M\sqrt{1-B^{2}}+  \nonumber \\
&&+\int dx\left\{ \frac{1}{2}\left( \frac{\partial \chi }{\partial x}\right)
^{2}+\frac{\pi ^{2}}{2}+\frac{1}{2}U_{sG}^{\prime \prime }\left( \Phi
_{c}\right) \chi ^{2}+V\text{-independent higher orders in }\chi \right\} + 
\nonumber \\
&&+\left\{ V(x+X)U_{sG}\left( \Phi _{c}\right) +V\text{-dependent higher
orders in }\chi \right\} +  \label{hkhi} \\
&&+B\int dx\pi \frac{\partial \chi }{\partial x}+ap_{B}+BP  \nonumber
\end{eqnarray}

\noindent The first terms in (\ref{41}) and (\ref{400c}) combine to yield
the first term in (\ref{hkhi}) above. This latter term yields the frequency
exponential $\exp (-M\sqrt{1-B^{2}}t)$ in the soliton wavefunction and is
the relativistic generalization of the phase $\exp (+imB^{2}t/2)$ (with an
apparently wrong sign) derived in \cite{9} for a point particle.

We can now write down the part $H_{0}$ of the total Hamiltonian (\ref{hkhi})
quadratic in the fluctuations: 
\begin{equation}
H_{0}=M\sqrt{1-B^{2}}+\int dx\left\{ \frac{1}{2}\left( \frac{\partial \chi }{%
\partial x}\right) ^{2}+\frac{\pi ^{2}}{2}+\frac{1}{2}U_{sG}^{\prime \prime
}\left( \Phi _{c}\right) \chi ^{2}\right\} +B\int dx\pi \frac{\partial \chi 
}{\partial x}  \label{50}
\end{equation}

\noindent The equation of motion describing free fluctuations derivable from
the above is 
\begin{equation}
\frac{\partial ^{2}\chi }{\partial t^{2}}-\frac{\partial ^{2}\chi }{\partial
x^{2}}+U_{sG}^{\prime \prime }\left( \Phi _{c}\right) \chi -2B\frac{\partial
^{2}\chi }{\partial x\partial t}+B^{2}\frac{\partial ^{2}\chi }{\partial
x^{2}}=0  \label{51}
\end{equation}

We can now check, as promised, that constraint $C$ commutes with the
quadratic Hamiltonian $H_{0}$: 
\begin{equation}
\left[ C,H_{0}\right] =\int dx\Phi _{c}^{\prime }\left\{ \frac{\partial
^{2}\chi }{\partial x^{2}}-U_{sG}^{\prime \prime }\left( \Phi _{c}\right)
\chi +B\frac{\partial \pi }{\partial x}\right\} +\frac{B}{\sqrt{1-B^{2}}}%
\int dx\Phi _{c}^{\prime \prime }\left\{ \pi +B\frac{\partial \chi }{%
\partial x}\right\}  \label{51b}
\end{equation}

\noindent Integrate by parts so that the above expression becomes linear in $%
\chi $ and $\pi $. The terms linear in $\pi $ cancel immediately. The
bracket multiplying $\chi $ vanishes due to equation (\ref{4b}) obeyed by $%
\Phi _{c}$.

\section{Determination of the acceleration.}

Constraint (\ref{420}) must commute with the total Hamiltonian $H$ (\ref{32}%
) and this determines $a$: 
\begin{equation}
\left[ C,H_{c}\right] +B\left[ C,\int dx\Pi \frac{\partial \Phi }{\partial x}%
\right] +a\left[ C,p_{B}\right] =0  \label{451a}
\end{equation}

\noindent An exact evaluation involves the fluctuating field in a somewhat
cumbersome way. However in the first instance we are interested in the
effect of the perturbing potential only (and disregard fluctuations). We use
(\ref{35}), (\ref{36}), take commutators (\ref{37}), (\ref{38}) into account
and keep terms proportional to $V$ only: 
\begin{equation}
Ma\frac{d}{dB}\left( \frac{B}{\sqrt{1-B^{2}}}\right) =\int dx\Phi
_{c}^{^{\prime }}\left( \frac{x}{\sqrt{1-B^{2}}}\right)
V(x+X)U_{sG}^{^{\prime }}(\Phi _{c})  \label{451}
\end{equation}

\noindent or 
\begin{equation}
M\frac{d}{dt}\left( \frac{B}{\sqrt{1-B^{2}}}\right) =\int dx\Phi
_{c}^{^{\prime }}\left( \frac{x}{\sqrt{1-B^{2}}}\right)
V(x+X)U_{sG}^{^{\prime }}(\Phi _{c})  \label{452}
\end{equation}

\noindent The left hand side is the relativistic expression for the force.
The right hand side depends on $X$, hence (\ref{452}) cannot be readily
integrated. The RHS of (\ref{452}) can be written 
\[
-\sin \left( k_{0}X\right) \int dx\Phi _{c}^{\prime }\left( \frac{x}{\sqrt{%
1-B^{2}}}\right) \Phi _{c}^{\prime \prime }\left( \frac{x}{\sqrt{1-B^{2}}}%
\right) \sin \left( k_{0}x\right) 
\]

\noindent where we used in turn (\ref{4b}) and (\ref{v}), expanded $V$ as a
sum of cosines and sines and kept the sine part (the cosine part vanishing
by parity). Rescaling 
\begin{equation}
M\frac{d}{dt}\left( \frac{B}{\sqrt{1-B^{2}}}\right) =-\sqrt{1-B^{2}}\sin
\left( k_{0}X\right) \int_{-\infty }^{\infty }dz\Phi _{c}^{\prime }(z)\Phi
_{c}^{\prime \prime }(z)\sin \left( k_{0}z\sqrt{1-B^{2}}\right)  \label{a2}
\end{equation}

\noindent Integrate the LHS\ by parts to get it in the form 
\[
\frac{1}{2}k_{0}\left( 1-B^{2}\right) \sin \left( k_{0}X\right)
\int_{-\infty }^{\infty }dz\left( \Phi _{c}^{\prime }(z)\right) ^{2}\cos
\left( k_{0}z\sqrt{1-B^{2}}\right) 
\]

\noindent and using (\ref{zero}) 
\[
2k_{0}\left( 1-B^{2}\right) \sin \left( k_{0}X\right) \int_{-\infty
}^{\infty }dz\frac{\cos \left( k_{0}z\sqrt{1-B^{2}}\right) }{\cosh ^{2}z} 
\]

\noindent The integral is standard and we get 
\begin{equation}
M\frac{d}{dt}\left( \frac{B}{\sqrt{1-B^{2}}}\right) =\frac{2\pi
k_{0}^{2}\left( 1-B^{2}\right) ^{3/2}}{\sinh \left( 
%TCIMACRO{\dfrac{\pi k_{0}\sqrt{1-B^{2}}}{2} }
%BeginExpansion
{\displaystyle {\pi k_{0}\sqrt{1-B^{2}} \over 2}}
%EndExpansion
\right) }\sin \left( k_{0}X\right)  \label{a3}
\end{equation}

The above expression connects relativistic acceleration and position.

\section{The normal modes.}

\subsection{The travelling modes.}

\subsubsection{The (t,x) frame.}

Solutions to (\ref{51}) are of the form 
\begin{equation}
A(B,K,x)\exp (-i\Omega t+iKx)  \label{52}
\end{equation}

\noindent \noindent and we set 
\[
g(B,K,x)=A(B,K,x)\exp (iKx) 
\]

\noindent To determine the relation between $\Omega $ and $K$ we observe
that away from the kink the modulating factor $A$ reduces to a constant;
then (\ref{51}) yields 
\begin{equation}
\Omega =-BK+\sqrt{K^{2}+\mu ^{2}}  \label{60}
\end{equation}

\noindent Note that $K$ caries a sign depending on the direction of
incidence. Also notice that $\Omega $ is positive (at least as long as the
kink does not move at superluminal speeds). Relation (\ref{60}) will also be
derived in the next paragraph by transforming from the inertial frame
instantaneously moving with the kink. The explicit form of $A(B,K,x)$ can be
deduced by struggling with (\ref{51}). We find it easier to deduce it in the
next paragraph via a Lorentz transformation.

It is clear from (\ref{51}) that due to the cross term in the derivatives
the $g$s are not orthogonal. However we can still proceed and introduce
creation and annihilation operators corresponding to the wavefunctions (\ref
{52}), the procedure now being somewhat more complicated than in standard
free field theory. We define (the index in $a^{\dagger }$ denoting the
direction of incidence being suppressed) 
\begin{equation}
a^{\dagger }(B,\Omega ,t)=\frac{1}{\sqrt{4\pi }}\int dxg^{*}(B,K,x)\left\{ 
\sqrt{\Omega }\chi -\frac{B}{i\sqrt{\Omega }}\frac{\partial \chi }{\partial x%
}+\frac{\pi }{i\sqrt{\Omega }}\right\}  \label{op1}
\end{equation}

\noindent with a corresponding expression for the complex conjugate. For $%
B=0 $ this reduces to the usual expression. To check that $a^{\dagger
}(B,\Omega ,t)$ satisfies the usual sinusoidal time evolution equation under
the influence of the quadratic Hamiltonian $H_{0}$ rewrite (\ref{51}) in the
form 
\begin{equation}
(1-B^{2})\frac{d^{2}g}{dx^{2}}+U_{sG}^{\prime \prime }\left( \Phi
_{c}\right) g+2iB\Omega \frac{dg}{dx}=-\Omega ^{2}g  \label{op2}
\end{equation}

\noindent Commute the RHS\ of (\ref{op1}) with $H_{0}$ (\ref{50}), integrate
by parts and use (\ref{op2}) to get in a straightforward manner 
\begin{equation}
\frac{d}{dt}a^{\dagger }(B,\Omega ,t)=i\Omega a^{\dagger }(B,\Omega ,t)
\label{op3}
\end{equation}

We shall later need the bracket $\left[ a^{\dagger }(B,\Omega ,t),\text{ }%
p_{B}\right] .$ To this end use the definition (\ref{op1}) and (\ref{37}), (%
\ref{38}) to get 
\begin{equation}
\left[ a^{\dagger }(B,\Omega ,t),\text{ }p_{B}\right] =-\frac{i}{\sqrt{4\pi
\Omega }}\frac{d}{dB}\left( \frac{B}{\sqrt{1-B^{2}}}\right) 
%TCIMACRO{\dint }
%BeginExpansion
\displaystyle \int 
%EndExpansion
dxg^{*}(B,K,x)\Phi _{c}^{\prime }\left( \frac{x}{\sqrt{1-B^{2}}}\right)
\label{op4a}
\end{equation}

\noindent with the complex conjugate for $a(B,\Omega ,t).$ Notice that in
the nonrelativistic version the integral in the RHS of the above equation
vanishes due to the orthogonality of the eigenfunctions of (\ref{5}).

We expand the fluctuation field in the form 
\begin{equation}
\chi (x,t)=\chi _{z}(x,t)+\frac{1}{\sqrt{2\pi }}\int \frac{d\Omega }{\sqrt{%
\Omega }}\left\{ a^{\dagger }(B,\Omega ,t)g(B,K,x)+a(B,\Omega
,t)g^{*}(B,K,x)\right\}  \label{op5}
\end{equation}

\noindent The above relation $defines$ the field $\chi _{z}(x,t)$ and is the
relativistic analog of (\ref{16}) (with the attendant complications of
non-orthogonality). To write the corresponding expansion for $\pi (x,t)$ we
are inspired (i) by the relation 
\begin{equation}
\frac{\partial \chi }{\partial t}=\pi +B\frac{\partial \chi }{\partial x}
\label{op6}
\end{equation}

\noindent derived from $H,$ (ii) by the fact that the expansion should be
valid in the special case of sinusoidal time evolution which holds when we
neglect effects due to the potential or to terms of degree higher than
quadratic in $\chi $. Thus 
\begin{eqnarray}
\pi (x,t) &=&\pi _{z}(x,t)+\frac{i}{\sqrt{2\pi }}\int d\Omega \left\{ \sqrt{%
\Omega }a^{\dagger }(B,\Omega ,t)g(B,K,x)-\sqrt{\Omega }a(B,\Omega
,t)g^{*}(B,K,x)\right\} +  \nonumber \\
&&+\frac{1}{\sqrt{2\pi }}\int \frac{d\Omega }{\sqrt{\Omega }}\left\{
a^{\dagger }(B,\Omega ,t)\frac{\partial g(B,K,x)}{\partial x}+a(B,\Omega ,t)%
\frac{\partial g^{*}(B,K,x)}{\partial x}\right\}  \label{op7}
\end{eqnarray}

\noindent Again this is the definition of $\pi _{z}(x,t).$ We take the
fields $\chi _{z}(x,t),\pi _{z}(x,t)$ to commute with the set of the $a$s
and $a^{\dagger }$s. The bracket $\left[ \chi _{z}(x,t),\pi _{z}(x^{\prime
},t)\right] $ is non-trivial in the nonrelativistic case as well. It equals 
\[
\frac{1}{M}\Phi _{c}^{\prime }(x)\Phi _{c}^{\prime }(x^{\prime }) 
\]

\noindent as can be seen from the zero mode part of (\ref{16}), (\ref{17}).

\subsubsection{The comoving frame.}

In the absence of interactions the kink moves at uniform velocity. Denote
quantities pertaining to the comoving frame by ($\sim $) and the lab frame
coordinates by prime. Consider a travelling wave of the form (\ref{14}), (%
\ref{14b}) 
\begin{equation}
\exp \left( -i\tilde{\omega}\tilde{t}+i\tilde{p}\tilde{x}\right) \tilde{A}%
\left( \tilde{p},\tilde{x}\right)  \label{65}
\end{equation}

\noindent The phase appearing in (\ref{65}) is written in terms of primed
coordinates 
\begin{equation}
-\tilde{\omega}\frac{t^{\prime }-Bx^{\prime }}{\sqrt{1-B^{2}}}+\tilde{p}%
\frac{x^{\prime }-Bt^{\prime }}{\sqrt{1-B^{2}}}  \label{62a}
\end{equation}

\noindent Use (\ref{21}) for uniform $B$ and rearrange 
\begin{equation}
-\tilde{\omega}\sqrt{1-B^{2}}t+\frac{\tilde{p}+B\tilde{\omega}}{\sqrt{1-B^{2}%
}}x  \label{62b}
\end{equation}

\noindent Compare with (\ref{52}) to identify 
\begin{equation}
\Omega =\tilde{\omega}\sqrt{1-B^{2}}{\ , }K=\frac{\tilde{p}+B\tilde{\omega}}{%
\sqrt{1-B^{2}}}  \label{62c}
\end{equation}

\noindent Thus $(\Omega ,K)$ is a rather peculiar pair: $K$ coincides with
the wavevector in the lab frame whereas $\Omega $ is connected to the
frequency in the comoving frame. Manipulation of $\Omega ,K$ as given above
yields again (\ref{60}). The Lorentz transformation together with (\ref{21})
yield the useful relation 
\begin{equation}
\tilde{x}=\frac{x^{\prime }-Bt^{\prime }}{\sqrt{1-B^{2}}}\Rightarrow \tilde{x%
}=\frac{x}{\sqrt{1-B^{2}}}  \label{67}
\end{equation}

\noindent The same Lorentz transformation yields 
\begin{equation}
\tilde{p}=\frac{K-B\sqrt{K^{2}+\mu ^{2}}}{\sqrt{1-B^{2}}}  \label{66}
\end{equation}

\noindent (Recall that we our choice of units $\mu =1$.) Thus the mode (\ref
{65}) is expressed in the $(t,x)$ frame in the form 
\begin{equation}
\exp (-i\Omega t)g(B,K,x)=\exp (-i\Omega t+iKx)\tilde{A}\left( \tilde{p}(K),%
\frac{x}{\sqrt{1-B^{2}}}\right)  \label{66b}
\end{equation}

\noindent Hence (\ref{66b}) provides the explicit form of the solutions (\ref
{52}) with $\tilde{A}$ taken from (\ref{14}), (\ref{14b}) and $\tilde{p}$
from (\ref{66}). In that sense it is hardly a mystery why operators $%
a^{\dagger }(B,\Omega ,t)$, $a(B,\Omega ,t)$ diagonalize the total
Hamiltonian: from the point of view of the comoving observer they correspond
to the usual (orthogonal) modes (\ref{14}), (\ref{14b}) of frequency $\Omega
/\sqrt{1-B^{2}}$.

\subsection{The zero mode.}

Consider an observer moving at velocity $B$ and suppose that the zero mode (%
\ref{30}) is excited, i.e. 
\[
\chi (\tilde{x},\tilde{t})=u\tilde{t}\Phi _{c}^{\prime }\left( \tilde{x}%
\right) 
\]

\noindent where $u$ is a small parameter (the velocity of the kink with
respect to the said observer). Transforming to the $\left( t,x\right) $
frame as above we get the same mode in the form 
\begin{equation}
\chi (x,t)=ut\sqrt{1-B^{2}}\Phi _{c}^{\prime }\left( \frac{x}{\sqrt{1-B^{2}}}%
\right) -u\frac{Bx}{\sqrt{1-B^{2}}}\Phi _{c}^{\prime }\left( \frac{x}{\sqrt{%
1-B^{2}}}\right)  \label{70}
\end{equation}

\noindent It can be readily shown that the above expression satisfies (\ref
{51}). This is precisely the form that $\chi _{z}(x,t)$ of (\ref{op5}) takes
in the free case. From (\ref{op6}), (\ref{op7}) and (\ref{70}) we can work
out the corresponding expression for $\pi _{z}(x,t).$

One can see constraint $C$ in a different light. It is easily shown that $C$
commutes with the $a$s and $a^{\dagger }$s and that in the special case of
free fluctuations it turns out to be proportional to $u$ of (\ref{70}). In
other words as seen from an observer moving at velocity $B$, the small
velocity $u$ associated with the excitation of the zero mode vanishes if $C$
is to hold; this is precisely what one would expect from a kink moving at
velocity $B$.

\section{Radiation from an accelerated soliton.}

To calculate the emission of radiation at a wavevector $K$ (with respect to
the lab frame) we look at the commutator $\left[ a^{\dagger }(B,\Omega
,t),H\right] $ ($H$ being the total Hamiltonian). We write 
\begin{equation}
\left[ a^{\dagger }(B,\Omega ,t),H\right] =(I)+(II)+(III)  \label{71}
\end{equation}

$(I)$ is given by (\ref{op3}) (corresponding to free evolution). The
commutator in (\ref{71}) gets contributions from terms in $H$ that are of
order higher than quadratic in $\chi $. However we are only interested in
contributions of first order in $V$ (i.e. of order $\varepsilon $; cf (\ref
{v})). Contributions $(II)$ and $(III)$ come from commuting (\ref{op1}) with
the term in the second pair of braces in (\ref{hkhi}) linear in $\chi $ and
with the $p_{B}$ term respectively. 
\begin{eqnarray}
(II) &=&-%
%TCIMACRO{\dfrac{\varepsilon }{i\sqrt{4\pi \Omega }} }
%BeginExpansion
{\displaystyle {\varepsilon  \over i\sqrt{4\pi \Omega }}}
%EndExpansion
\int_{-\infty }^{\infty }dxg^{*}(B,k,x)V(x+X)U_{sG}^{\prime }\left( \Phi
_{c}\right) =  \label{72} \\
&=&-%
%TCIMACRO{\dfrac{\varepsilon }{i\sqrt{4\pi \Omega }} }
%BeginExpansion
{\displaystyle {\varepsilon  \over i\sqrt{4\pi \Omega }}}
%EndExpansion
\int_{-\infty }^{\infty }dxg^{*}(B,k,x)\Phi _{c}^{\prime \prime }\left( 
\frac{x}{\sqrt{1-B^{2}}}\right) \cos \left( k_{0}x+k_{0}X\right)  \nonumber
\end{eqnarray}

\noindent Recall that the $X$ has a non-trivial time dependence.

For $(III)$ we get from (\ref{op4a}) 
\begin{equation}
(III)=-a\frac{i}{\sqrt{4\pi \Omega }}\frac{d}{dB}\left( \frac{B}{\sqrt{%
1-B^{2}}}\right) 
%TCIMACRO{\dint }
%BeginExpansion
\displaystyle \int 
%EndExpansion
dxg^{*}(B,K,x)\Phi _{c}^{\prime }\left( \frac{x}{\sqrt{1-B^{2}}}\right)
\label{73}
\end{equation}

\noindent Given that $a=dB/dt$ the combination 
\[
a\frac{d}{dB}\left( \frac{B}{\sqrt{1-B^{2}}}\right) 
\]

\noindent amounts to $1/M$ times the right hand side of (\ref{a3}). Notice
that the RHS of (\ref{73}) does not vanish since the solutions of (\ref{351}%
) are not orthogonal. Thus 
\[
(III)=-\frac{i}{M\sqrt{4\pi \Omega }}\frac{2\pi k_{0}^{2}\left(
1-B^{2}\right) ^{3/2}}{\sinh \left( 
%TCIMACRO{\dfrac{\pi k_{0}\sqrt{1-B^{2}}}{2} }
%BeginExpansion
{\displaystyle {\pi k_{0}\sqrt{1-B^{2}} \over 2}}
%EndExpansion
\right) }\sin \left( k_{0}X\right) 
%TCIMACRO{\dint }
%BeginExpansion
\displaystyle \int 
%EndExpansion
dxg^{*}(B,K,x)\Phi _{c}^{\prime }\left( \frac{x}{\sqrt{1-B^{2}}}\right) 
\]

\section{Conclusion.}

This paper is meant to set the general framework for a treatment of the
collective coordinate of a relativistic soliton interacting with
fluctuations and an external potential. In a relativistic setting it proves
natural to treat the velocity $B$ as a dynamical variable. This artificial
increase in the number of degrees of freedom and the attendant gauge
invariance generated by the constraint (\ref{27}) are lifted by the
subsidiary condition that defines the position of the soliton. Let us stress
again that the formulae for the total momentum and energy of the system (\ref
{41}) and (\ref{400c}) include the correct relativistic expressions for the
soliton momentum and energy. It is equally reassuring that the left hand
sides of (\ref{452}) and (\ref{a3}) feature the time derivative of the
relativistic momentum.

Concerning emission of radiation we observe that it has two physical
origins. One is the interaction of the soliton with the fluctuations. This
comes about when we commute the $\pi $ dependent term in (\ref{op1}) with
the $V$ independent cubic and quartic terms in $\chi $ in the total
Hamiltonian. This contribution was disregarded in the present work only
because we were interested in the $V$ dependence. The acceleration also
receives contributions from terms in the Hamiltonian depending on both $V$
and $\chi ,$ reflecting the fact that an accelerated soliton radiates; to
lowest order we obtain ($II)$ calculated above. Finally observe that
contribution $(III)$ stems from the $p_{B}$ term in the Hamiltonian, whose
existence is required by the need of a relativistic treatment.

\section{Acknowledgments.}

The author is indebted to Professor Gabriel Barton for his comments and to
Professor Grigori Volovik for discussions. He also wishes to thank the Low
Temperature Laboratory of Helsinki University of Technology for its
hospitality and EU\ TMR programme ERBFMGECT980122 for support.

\end{document}